\documentclass[draft]{agujournal2019}
\usepackage{url} 
\usepackage{lineno}
\usepackage[inline]{trackchanges} 
\usepackage{soul}
\draftfalse

\journalname{JGR: Space Physics}

\begin{document}

%
%

\title{Correlation between bandwidth and frequency of plasmaspheric hiss uncovered with unsupervised machine learning}

\authors{Daniel Vech\affil{1}, David M. Malaspina\affil{1,2}, Alexander Drozdov\affil{3} and Anthony Saikin\affil{3}}
\affiliation{1}{Laboratory for Atmospheric and Space Physics, University of Colorado Boulder, Boulder, CO, USA}
\affiliation{2}{Department of Astrophysical and Planetary Sciences, University of Colorado Boulder, Boulder, Colorado, USA}
\affiliation{3}{Department of Earth, Planetary and Space Sciences, University of California, Los Angeles, USA}

\correspondingauthor{Daniel Vech}{daniel.vech@lasp.colorado.edu}




\begin{keypoints}
\item Unsupervised machine learning is used to categorize hiss power spectra of the electric field from Van Allen Probes
\item From the pre-noon sector towards the afternoon sector the hiss frequency decreases while the bandwidth increases
\item We discuss possible source mechanisms that are consistent with the spatial distribution of the hiss waves

\end{keypoints}

%
%

%
%


\begin{abstract}

Previous statistical studies of plasmaspheric hiss investigated the averaged shape of the magnetic field power spectra at various points in the magnetosphere. However, this approach does not consider the fact that very diverse spectral shapes exist at a given L-shell and magnetic local time. Averaging the data together means that important features of the spectral shapes are lost. In this paper, we use an unsupervised machine learning technique to categorize plasmaspheric hiss. In contrast to the previous studies, this technique allows us to identify power spectra that have "similar" shapes and study their spatial distribution without averaging together vastly different spectral shapes. We show that strong negative correlations exist between the hiss frequency and bandwidth, which suggests that the observed patterns are consistent with in situ wave growth.

\end{abstract}

\section*{Plain Language Summary}
[ enter your Plain Language Summary here or delete this section]

%
%

%


%
%
%
%

\section{Introduction}

Plasmaspheric hiss is an incoherent, electromagnetic whistler mode wave that has a frequency range of 20 Hz to a few kHz. Hiss plays a major role in the scattering of high energy electrons and creating the slot region between the inner and outer radiation belts \cite{thorne1973plasmaspheric}. Observations by \citeA{li2013unusual} suggested that hiss can be split into two components: low ($<$150 Hz) and high ($>$150 Hz) frequency hiss waves. Recent analysis by \citeA{malaspina2017statistical} showed that the low and high frequency waves are statistically distinct populations. Two major differences are that low frequency hiss reaches peak amplitudes near 15 hours magnetic local time (MLT), while it is approximately 12 MLT for the high frequency one. Also, low frequency hiss is localized close to the plasmapause while high frequency hiss can be observed significantly farther earthward. Understanding the properties of low frequency hiss is particularly important because it can resonate with higher energy electrons than the high frequency part, therefore it may have a larger impact on the radiation belt dynamics by scattering electrons out of trapped orbits and into the atmosphere \cite{li2013unusual}.

Previous studies of hiss used either case studies, which were manually identified ($<$100 wave events, \citeA{chen2014generation, ni2014resonant,li2013unusual}) or analyzed the statistical shape of the magnetic field power spectra by averaging thousands of hours of data as a function of L-shell, MLT, and distance from the plasmasphere \cite{malaspina2017statistical,meredith2018global,meredith2021statistical}. While a statistical approach is necessary to obtain robust results of the spatial distribution of hiss wave activity, this method has a major disadvantage: a wide range of spectral shapes (waves with different center frequencies and bandwidths) co-exist in a given spatial domain (e.g. MLT and L-shell bin and geomagnetic activity), therefore averaging them together means that important details about the wave activity are lost. This is particularly problematic for accurate inclusion of hiss wave population in predictive models of inner magnetosphere plasma dynamics since the statistical spectral shape might be significantly distorted due to the averaging \cite<e.g.>[]{chen2012modeling}.


In this paper, we use an unsupervised machine learning technique called Self-Organizing Map (SOM) to identify and categorize plasmaspheric hiss. This technique sorts electric field power spectra into nodes where power spectra belonging to the same node have similar properties: they all display wave activity at approximately the same frequency with similar power spectra density, and bandwidth.  This method has the advantage that large data sets ($>$1 million electric field power spectra) can be analyzed without spatial averaging of a broad range of spectral shapes, therefore the key properties of hiss (frequency, bandwidth and power spectra density) can be derived more accurately. We investigate the spatial distribution of various electric field power spectra shapes in the plasmasphere and show that from 10 hours MLT to 14 hours MLT the hiss frequency increases while the bandwidth decreases. We discuss the possible mechanisms that may explain the origin of the low and high frequency hiss.

\section{Data Preparation and Methodology}

We use the Van Allen Probes data sets previously analyzed by \citeA{malaspina2017statistical}, which is based on measurements from the Electric Fields and Waves (EFW) instrument \cite{wygant2013electric} and the Electric and Magnetic Field Instrument Suite and Integrated Science (EMFISIS) instrument suite \cite{kletzing2013electric}. Data outside the plasmasphere ($n_e<$50 cm$^{-3}$) and data recorded during spacecraft charging events, eclipses, thruster firings or EFW bias sweeps were excluded from the analysis. For details of the data cleaning see \cite{malaspina2017statistical}. The spin and axial electric field power spectra were measured onboard for 0.5 seconds out of 6 second intervals on a logarithmically spaced frequency grid with 50 elements between 2 and 2000 Hz. As opposed to several previous studies \cite{malaspina2017statistical,meredith2018global,meredith2021statistical}, we use the electric field data to analyze the spectral properties of hiss waves. We suggest that the electric field instrument is more appropriate for the analysis of low frequency hiss compared to the search coil magnetometer due to relatively lower noise floor compared to plasma wave signals at frequencies $<$200 Hz \cite{wygant2013electric}. The relatively high noise floor of the search coil magnetometer at frequencies $<$200 Hz leads to the systematic overestimation of the power spectra density in that frequency range, which was discussed by \citeA{malaspina2017statistical}.

The combined (Probe A and B) data set includes over 24.6 million power spectra, which significantly exceeds the size that could be processed with our computational resources to train a Self-Organizing Map. Therefore, we restrict our study to 250 days of randomly selected data from Probe A (2.1 million power spectra).

The presence of magnetosonic waves can distort the magnetic and electric field power spectra. We use the following filtering method to eliminate them from our analysis: the compressibility ($|\delta B_{||}|/|\delta B_{total}|$, for details see \citeA{malaspina2016distribution}) of magnetic fluctuations is calculated in 50 frequency bins between 20-2000 Hz. We omit all of those electric field power spectra where the corresponding magnetic compressibility spectra had more than 6 frequency bins with $|\delta B_{||}|/|\delta B_{total}|>$0.6. In total 245,000 power spectra (from the initial 2.1 million) were excluded from the hiss analysis due to this criteria.

We use the machine learning technique developed by \citeA{vech2021novel}, which was demonstrated with large data sets (182,000 power spectra in total) of fluxgate and search coil magnetic field data from the Magnetospheric Multiscale Mission. SOM is an unsupervised machine learning technique that consists of a two-dimensional grid of nodes where the number of nodes is typically between a few dozens and a few hundreds; in our study we use 100 nodes. The goal of the training process is to assign each input vector (i.e. power spectra) to a node while ensuring that "similar" input vectors are assigned to the same or neighboring nodes while "dissimilar" input vectors are assigned to nodes far from each other. The similarity between input vectors can be quantified by a variety of metrics, here we use the Euclidian distance: $d(q,p)=\sqrt{\sum(q_i - p_i})^2$ where $q$ and $p$ correspond to a pair of power spectra and $d$ quantifies their similarity.

Since the power spectra density of the electric field fluctuations is highly variable, the power spectra have to be normalized before the SOM training process. We shift (in power spectra density) the electric field power spectra with a constant factor so they are all set to 0 (in logarmithmic space) (V/m)$^2$/Hz at 20 Hz. This normalization means that the differences (i.e. the value of $d(q,p)$) between the power spectra are determined by differences in the high frequency wave activity and therefore the effect of low ($<$20 Hz) frequency fluctuations is eliminated.

We used the procedure described in \citeA{vech2021novel} and trained the SOM with a 10x10 grid of nodes. The input matrix has 2.1 million rows (number of power spectra) and 50 columns (number of frequency bins) based on the electric field power spectra from the spin axis sensors. We do not use the axial electric field for the SOM training because this data product is affected by artifacts due to the fact that the voltage sensor is periodically in the shadow of the spacecraft \cite{kletzing2013electric}. The training process was repeated 10 times (500 iterations each time) and we found that approximately 0.1\% of the input vectors were assigned to different nodes suggesting that the trained model converged to a steady state and
the node-assignment variation between iterations became small.

In order to illustrate the power spectra assigned to a node, we plot the average power spectra for three nodes in Figure 1. The error bars correspond to the standard deviation of the power spectra density for each frequency bin. Figure 1a shows an example for a node that has no significant wave activity in the range of 20 Hz to 2000 Hz. Figure 1b shows a node that has high frequency hiss due to the fact the peak power spectra density (frequency corresponding to the "bump" in the spectra) is approximately at 600 Hz. Finally, Figure 1c shows an example of a node with low frequency hiss due to the fact that the enhanced wave activity extends well below 150 Hz.

\begin{figure}
    \centering\includegraphics[width=1\linewidth]{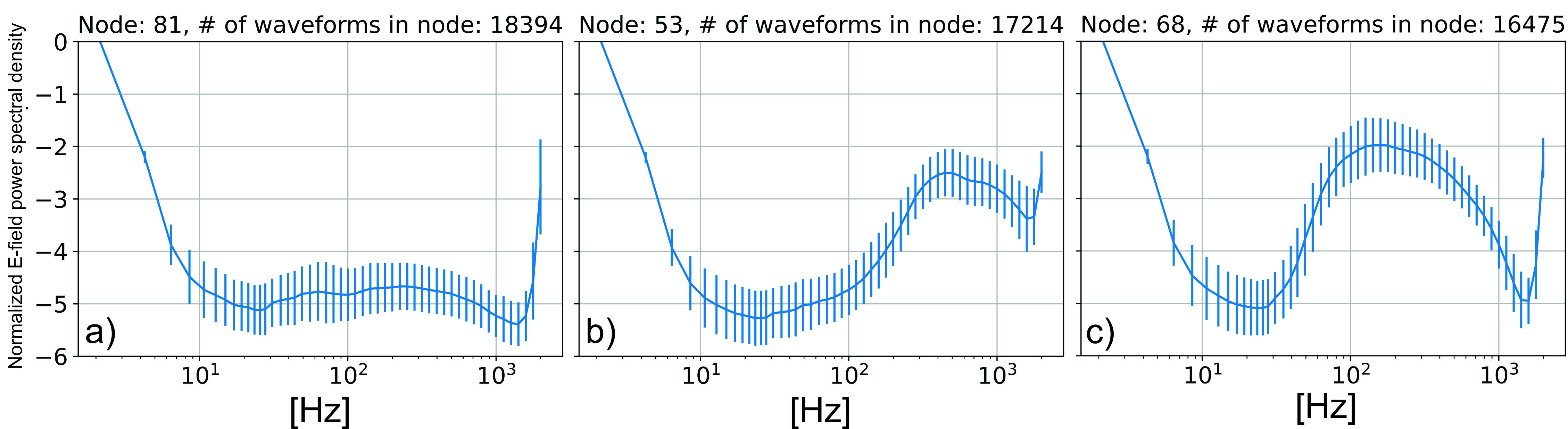}
\caption{Three examples of the nodes a) without significant wave activity, b) with high frequency hiss and c) with low frequency hiss. The line plots correspond to the average of all power spectra assigned to each of the three nodes. The error bars correspond to the standard deviation of the E-field power spectra density in each frequency bin.}
  \label{fig:1}
\end{figure}

We use the following method to identify nodes that display "significant" wave activity. We integrate the node averaged power spectra (such as Figure 1a,b and c) from 20 Hz to 1000 Hz (P) and split the data into two groups as P$<$1.57 $V/m$ (34 nodes) and P$>$1.57 $V/m$ (66 nodes). This empirical threshold was determined after manual inspection of all the 100 nodes and was found to be an adequate point to split the nodes into "no waves" and "waves" categories.

For those 66 nodes with wave activity, we define the wave frequency and bandwidth with the following metrics. For each node, we identify the frequency of the inflection point (corresponding to the peak wave power spectra density) in the power spectra as the "wave frequency". For example, this is approximately 455 Hz in Figure 1b and 130 Hz in Figure 1c. The bandwidth is measured as the ratio of frequencies (below and above the "wave frequency") where the power spectra density drops (from the peak) by the factor of 1/e measured in logarithmic space. For example, this is a factor of 2.55 in Figure 1b and 1.4 in Figure 1c, respectively.



\section{Spatial distribution of low and high frequency hiss}

In this Section, we investigate the spatial distribution of the observed hiss waves. We plot the distribution of hiss wave characteristics in a grid with with 36 angular (MLT from 0 to 24 hr) and 7 radial (L-shell from 0 to 7 Earth radii) bins.

\begin{figure}
    \centering\includegraphics[width=1\linewidth]{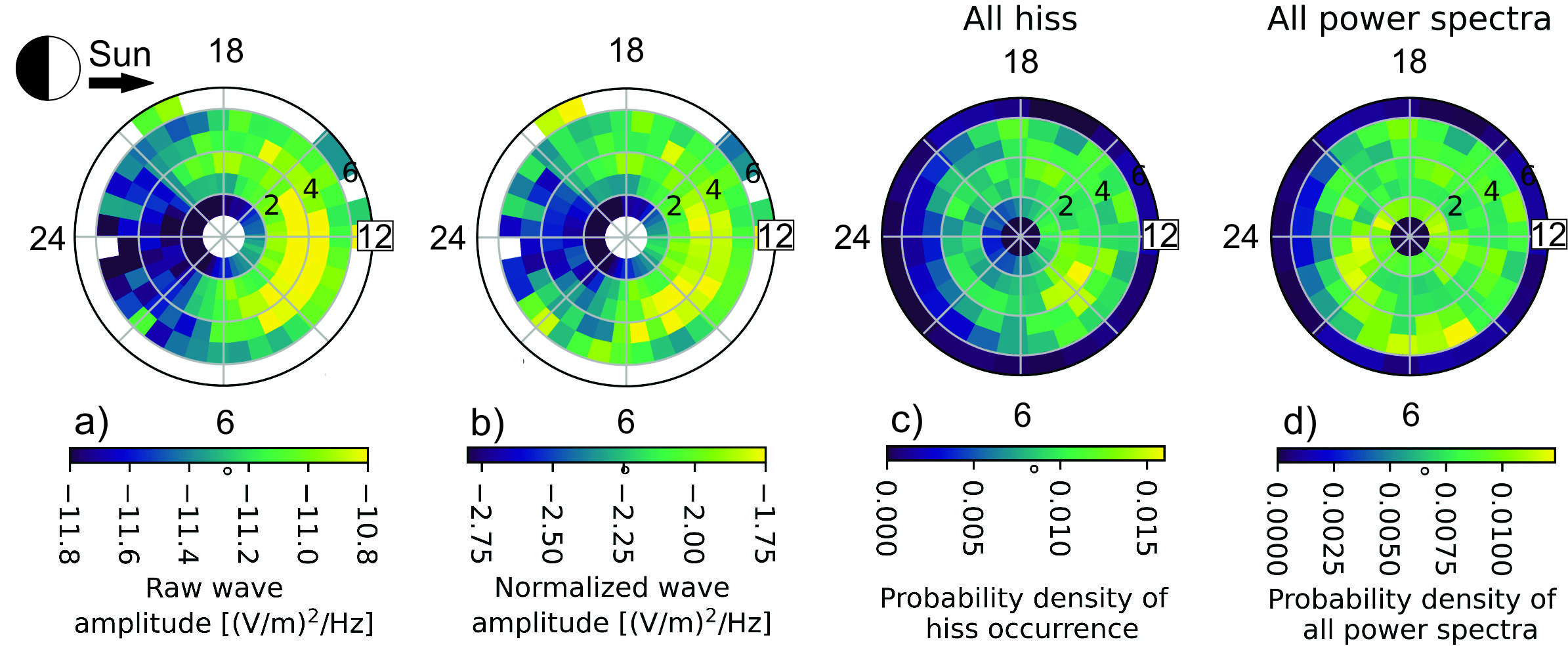}
\caption{Median hiss wave power spectra density in each bin for the normalized (a) and raw data (b) on log scale. (c) Spatial distribution of the power spectra with hiss wave activity. (d) Spatial distribution of all power spectra used in our study regardless the wave activity.}
  \label{fig:2}
\end{figure}

First we plot the hiss wave power spectra density at the wave frequency for the selected 66 nodes incorporating 1.1 million power spectra in the L-shell vs. MLT grid in Figure 2a and b where the color code corresponds to the median power spectra density in each L-shell vs. MLT bin on log scale. The distribution of this normalized power spectra density is shown in Figure 2a. In Figure 2b, we investigate the distribution of the "raw power spectra density" that refers to the power spectra density without normalization. The two panels display different features of the data set: Figure 2a essentially shows the hiss power spectra density with respect to the power spectra density at 20 Hz (i.e a relative power spectra density), in contrast Figure 2b show the "absolute value" of the hiss power spectra density. Both the normalized and raw power spectra density show some bias toward the pre-noon sector, which is consistent with the findings of previous statistical studies such as \citeA{meredith2018global, meredith2021statistical}.

\begin{figure}
    \centering\includegraphics[width=1\linewidth]{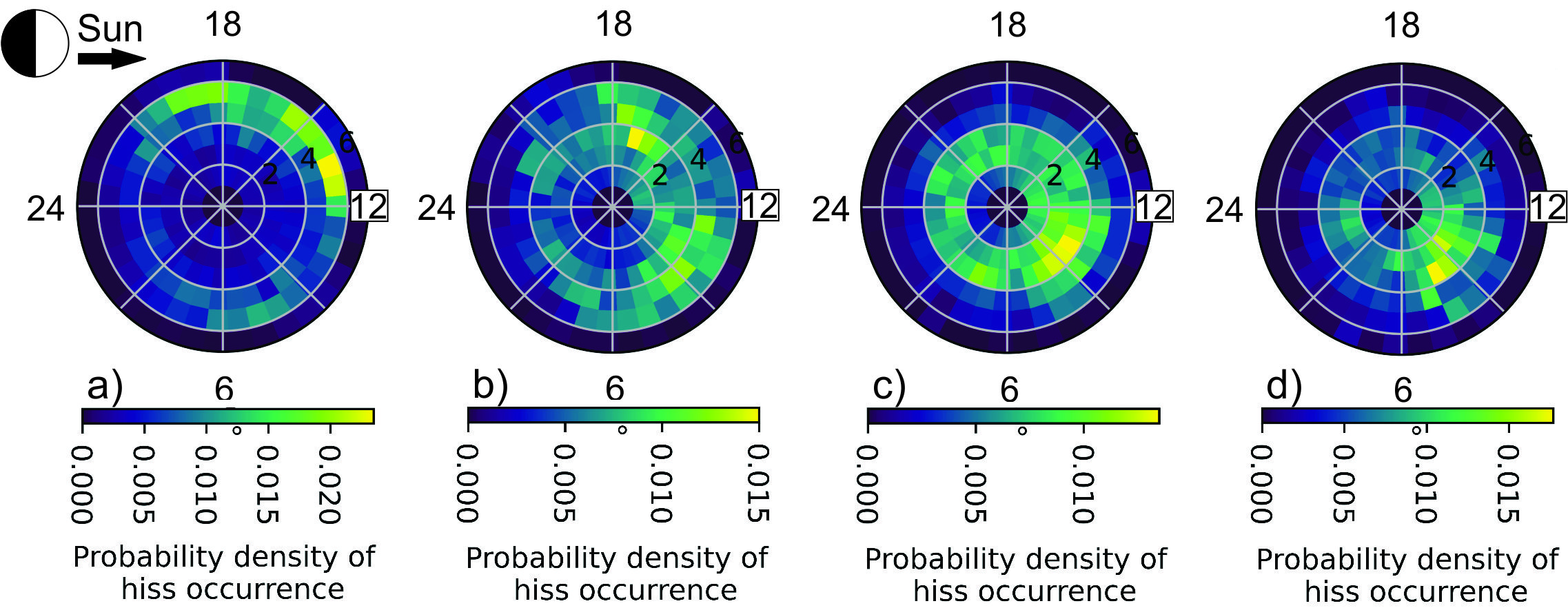}
\caption{Spatial distribution of hiss with peak frequency a) f $<$ 194 Hz, b) 194 $<$ f $<$ 252 Hz, c) 252 $<$ f $<$ 316 Hz, d) 316 Hz $<$ f, respectively.}
  \label{fig:3}
\end{figure}

\begin{figure}
    \centering\includegraphics[width=1\linewidth]{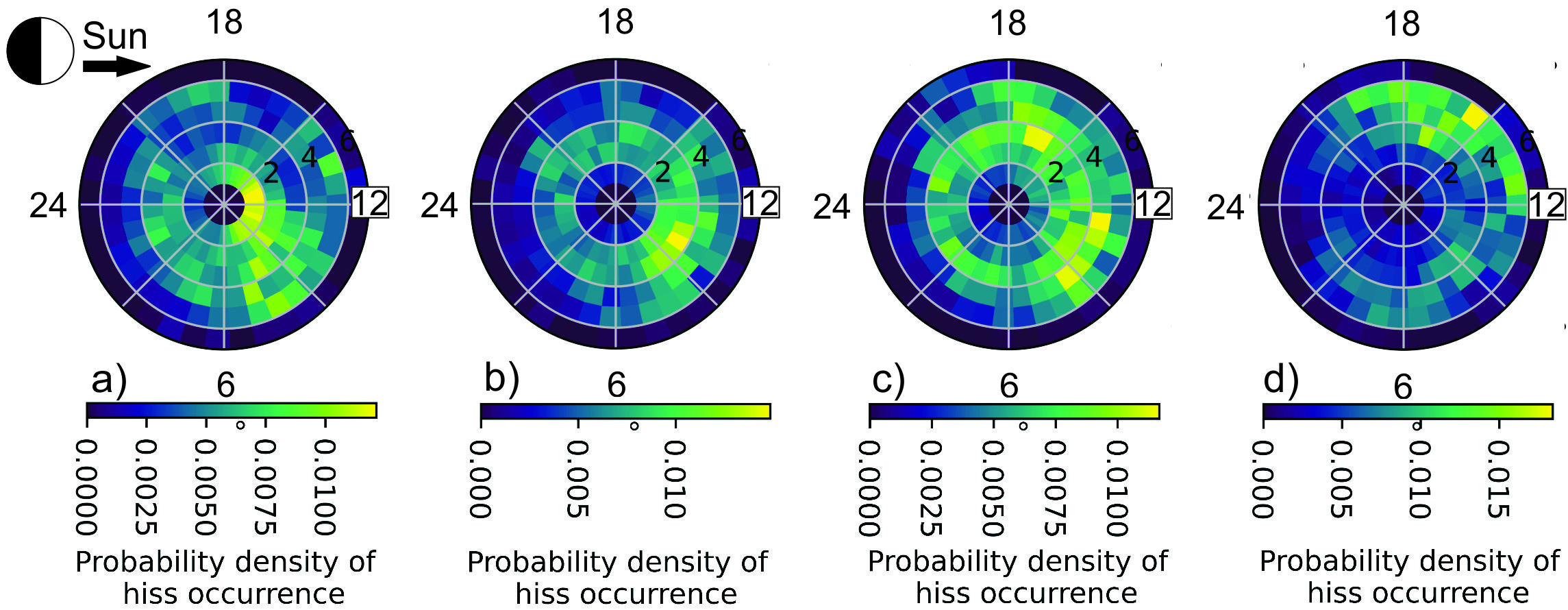}
\caption{Spatial distribution of hiss with bandwidth (B) in the range of a) B $<$ 2.05, b) 2.05 $<$ B $<$ 2.51, c) 2.51 $<$ B $<$ 2.83, d) B $>$ 2.83.}
  \label{fig:4}
\end{figure}

In Figure 2c we investigate the rate of occurrence of hiss by plotting the spatial distribution of the 1.1 million power spectra that were assigned to the 66 nodes with wave activity. The color code corresponds to the probability density in each bin, which is obtained by normalizing the count number in a given bin by the sum of all power spectra in the plot. Previous statistical studies such as \citeA{malaspina2017statistical,meredith2018global,meredith2021statistical} analyzed the distribution of power spectra amplitude (or integrated wave power) in the L-shell vs. MLT space, however, it is important to note that the maps presented in previous papers do not necessarily correspond to the rate of occurrence of hiss. A region with very large hiss amplitude does not necessarily coincide with the region where hiss occurs most frequently. Determination of the rate of occurrence requires classification of the power spectra (at the minimum two categories as "hiss" vs. "no hiss"), which was achieved with the SOM training process. Hiss occurrence rate is concentrated mostly in the pre-noon sector (Figure 3c), while the enhanced hiss amplitude extends to noon (Figure 2a,b).

In Figure 2d, we plot the spatial location of each of the 2.1 million power spectra (i.e. all power spectra, regardless the wave activity) that we used in our study. The distribution suggests that the increased rate of hiss occurrence in the pre-noon sector presented in Figure 2a,b and c are not due to imbalance in the spacecraft observations (i.e. having more observations from 9-12 MLT compared to the rest of the spatial locations).


Previously we determined the frequency and bandwidth of the nodes that displayed wave activity. Using these derived parameters, we create further sub-categories of the power spectra. First, we split the data (1.1 million power spectra assigned to the 66 nodes with hiss) into four groups based on the wave frequency. The frequency (f) thresholds are: 1) f $<$ 194 Hz, 2) 194 $<$ f $<$ 252 Hz, 3) 252 $<$ f $<$ 316 Hz, 4) 316 Hz $<$ f. These thresholds were determined as the 25, 50, 75 percentile values of the wave frequency data. We apply the same approach to the bandwidth (B) where the thresholds are 1) B $<$ 2.05, 2) 2.05 $<$ B $<$ 2.51, 3) 2.51 $<$ B $<$ 2.83, 4) B $>$ 2.83.

Figures 3 and 4 show the spatial distribution of hiss for each sub-category. First we compare Figure 3a and Figure 4d: the overlap between these plots suggests that hiss in the pre-noon sector is characterized by narrow bandwidth and high frequency. The comparison of Figure 3d and Figure 4a shows that in the afternoon sector the hiss has the lowest frequency and broadest bandwidth. Although there is some scattering in the intermediate categories of Figure 3b,c and Figure 4b,c, they are consistent with the pattern that the hiss bandwidth increases from pre-noon to afternoon while the frequency decreases.

\section{Generation mechanisms of low and high frequency hiss}

Historically, the two main leading theories of hiss growth were the 1) in situ growth due to unstable electron distributions \cite<e.g.>[]{church1983origin} and 2) wave injection due to terrestrial lightning strikes \cite<e.g.>[]{draganov1992magnetospherically}. However, more recent measurements show that both of these concepts are often inconsistent with data \cite<e.g.>[]{green2005origin}. 

\citeA{bortnik2008unexpected} offered an alternative explanation suggesting that chorus waves may be the source of hiss. This model could explain several features of hiss such as distribution in L-shell (pre-noon sector) and day-night asymmetry. A statistical study of the correlation between hiss and chorus waves found significant overlap (in MLT) between the two wave modes \cite{agapitov2018spatial}. However, the chorus origin is also a matter of considerable debate. For example, \citeA{hartley2019van} suggested that the wave-vector orientation of chorus waves in the pre-noon sector is not consistent with entering the plasmasphere and in the pre-noon sector chorus waves can explain only $\approx1\%$ of hiss wave power. The transition of chorus waves to hiss waves was found to be more significant near the plasmaspheric plume where large azimuthal density gradients exist. In that region $>80\%$ of the hiss wave power was explained by chorus waves.

A major difficulty for the chorus wave theory is to explain the "frequency jump" across the plasmasphere. Chorus waves are typically observed between 0.1-1 $\omega_{ce}$ (corresponding to approximately 2-7 kHz and $\omega_{ce}$ is electron cyclotron frequency) \cite{bortnik2008unexpected} while the the typical hiss frequency in the pre-noon sector is around 316-1000 Hz meaning the required frequency change is a factor of 2 to 22. Moving toward the dusk region, the hiss frequency decreases ($<$194 Hz, Figure 3), therefore the required frequency jump for chorus to transition into hiss could be as large as a factor of 10 to 40. In addition to this difficulty, the rate of occurrence of chorus waves drops significantly in the $>$12 MLT region compared to the pre-noon sector \cite{agapitov2018spatial}. Therefore, even if chorus waves can enter the plasmasphere in the afternoon sector, they can only explain a small fraction ($\approx 24\% $) of all hiss occurrence.


Recently several studies argued that electron injections may play an important role in hiss generation. For example, \citeA{shi2017systematic} conducted a statistical study of low frequency hiss and suggested that local amplification induced by electron injection events at higher (L$\approx$6) L-shell is a possible source of these waves. The statistical distribution of low frequency hiss (Figure 3a) is consistent with this idea. \citeA{hikishima2020particle}, proposed local generation of hiss through linear and nonlinear interactions of electromagnetic field fluctuations with anisotropic energetic electrons. \citeA{ratcliffe2017self} used particle-in-cell simulations to model whistler mode wave growth with a distinctly warm and hot electron populations. They found that the growth of whistler mode waves was split into upper and lower bands approximately around 0.5$\omega_{ce}$. They also found that the frequency gap sensitively depends on the temperature and anisotropy of each electron component.
\citeA{zhu2019triggered} suggested that low-frequency hiss consists of parallel and antiparallel Poynting fluxes, resulting from multiple reflections inside the plasmasphere.

We suggest that differences in the electron populations between the pre-noon and afternoon sectors might be able to split hiss waves into two bands in a similar fashion. There is some evidence for this idea in Figure 3b and c, which shows that the high and low frequency hiss are strongly separated, and there is a gap in the rate of hiss occurrence at around 12 MLT in Figure 3b. This suggests that the different electron distributions in the pre-noon and afternoon sectors may support hiss wave growth in two separate frequency bands. 





\section{Conclusion}

The traditional approach for studying plasmaspheric hiss is based on calculating spatial averages of the magnetic field power spectra. This technique has a major disadvantage since it does not take into account the diverse shapes of power spectra that occur in a given L-shell vs. MLT bin. In this paper, we used an unsupervised machine learning technique to categorize plasmaspheric hiss and studied the spatial distribution of the various spectral shapes without averaging together vastly different spectral shapes.

First, we categorized the power spectra as "hiss" vs. "no hiss" and studied the rate of occurrence of hiss in the L-shell vs. MLT space.  Secondly, we created eight sub-categories of hiss based on bandwidth and frequency. This sophisticated classification allowed us to understand the evolution of the spectral shapes from dawn to dusk. We showed that hiss at around 9 MLT have the narrowest bandwidth and highest frequency. The frequency gradually decreases toward dusk while the bandwidth broadens. 

We discussed possible mechanisms that may generate plasmaspheric hiss and pointed out some inconsistencies between our observations and the idea that hiss originates from chorus waves. To explain the obtained frequency and bandwidth correlation, we favor the in situ wave growth mechanism proposed by \citeA{ratcliffe2017self} due to the fact that it could naturally account for the observed two bands of hiss waves. Further work is needed to quantify the required temperature and anisotropy of the hot and warm electron populations that create hiss waves consistent with our observations.


Finally, the some current radiation belt models operate with simple assumptions such as constant hiss frequency and amplitude \cite<e.g.>[]{fok2011recent}. In our study we quantified the variability of the hiss spectral shapes. Based on the results we suggest that parameterizing the hiss with MLT dependent frequency and bandwidth is necessary for adequate inclusion of this wave mode in predictive models of high energy electron scattering.

\acknowledgments
D. V. was supported by NASA contract 80NSSC21K0454. D.M. was supported by NASA contract 80NSSC19K0305. The authors thank the Van Allen Probes team, especially the EFW and EMFISIS teams for their support. This work was funded by NASA Grant 80NSSC18K1034. All Van Allen Probes data used in this work are available from the EFW and EMFISIS team websites (which one can link to here:http://rbspgway.jhuapl.edu). 

\end{document}